\documentclass[pmlr,color]{jmlr}% new name PMLR (Proceedings of Machine Learning Research)

\usepackage{longtable}% for long tables

\usepackage{booktabs}
\usepackage[load-configurations=version-1]{siunitx} % newer version
\theorembodyfont{\upshape}
\theoremheaderfont{\scshape}
\theorempostheader{:}
\theoremsep{\newline}

\jmlrvolume{}
\jmlryear{2022}
\jmlrworkshop{PREPRINT}

\title[Masked Spectrogram Modeling using Masked Autoencoders]{Masked Spectrogram Modeling using Masked Autoencoders for Learning General-purpose Audio Representation}

  \author{\Name{Daisuke Niizumi} \Email{daisuke.niizumi.dt@hco.ntt.co.jp}\\
   \Name{Daiki Takeuchi} \Email{daiki.takeuchi.ux@hco.ntt.co.jp}\\
   \Name{Yasunori Ohishi} \Email{yasunori.ooishi.uk@hco.ntt.co.jp}\\
   \Name{Noboru Harada} \Email{noboru.harada.pv@hco.ntt.co.jp}\\
   \Name{Kunio Kashino} \Email{kunio.kashino.me@hco.ntt.co.jp}\\
   \addr NTT Communication Science Laboratories, NTT Corporation, Atsugi, Japan}

\editors{Joseph Turian, Björn W. Schuller, Dorien Herremans, Katrin Kirchhoff, Paola Garcia Perera, and Philippe Esling}

\begin{document}

\maketitle

\begin{abstract}
Recent general-purpose audio representations show state-of-the-art performance on various audio tasks.
These representations are pre-trained by self-supervised learning methods that create training signals from the input. For example, typical audio contrastive learning uses temporal relationships among input sounds to create training signals, whereas some methods use a difference among input views created by data augmentations.
However, these training signals do not provide information derived from the intact input sound, which we think is suboptimal for learning representation that describes the input as it is.

In this paper, we seek to learn audio representations from the input itself as supervision using a pretext task of auto-encoding of masked spectrogram patches, Masked Spectrogram Modeling (MSM, a variant of Masked Image Modeling applied to audio spectrogram).
To implement MSM, we use Masked Autoencoders (MAE), an image self-supervised learning method.
MAE learns to efficiently encode the small number of visible patches into latent representations to carry essential information for reconstructing a large number of masked patches.
While training, MAE minimizes the reconstruction error, which uses the input as training signal, consequently achieving our goal.

We conducted experiments on our MSM using MAE (MSM-MAE) models under the evaluation benchmark of the HEAR 2021 NeurIPS Challenge.
Our MSM-MAE models outperformed the HEAR 2021 Challenge results on seven out of 15 tasks (e.g., accuracies of 73.4\% on CREMA-D and 85.8\% on LibriCount), while showing top performance on other tasks where specialized models perform better.
We also investigate how the design choices of MSM-MAE impact the performance and conduct qualitative analysis of visualization outcomes to gain an understanding of learned representations.
We have made our code available online for further improvements and applications of the MSM framework.\footnote{\url{https://github.com/nttcslab/msm-mae}}
\end{abstract}

\section{Introduction}\label{sec:intro}
With the recent progress of audio representation learning, general-purpose audio representations have shown good performance in various audio tasks \citep{saeed2020cola,niizumi2021byol-a,wang2021universal}.
While previous supervised learning methods \citep{hershey2017cnn,kong2020panns,koutini2021passt} learn to discriminate labels, these general-purpose audio representations are pre-trained by self-supervised learning methods that do not rely on labels.

\begin{figure}[htbp]
  \centering
  \includegraphics[width=0.6\linewidth]
  {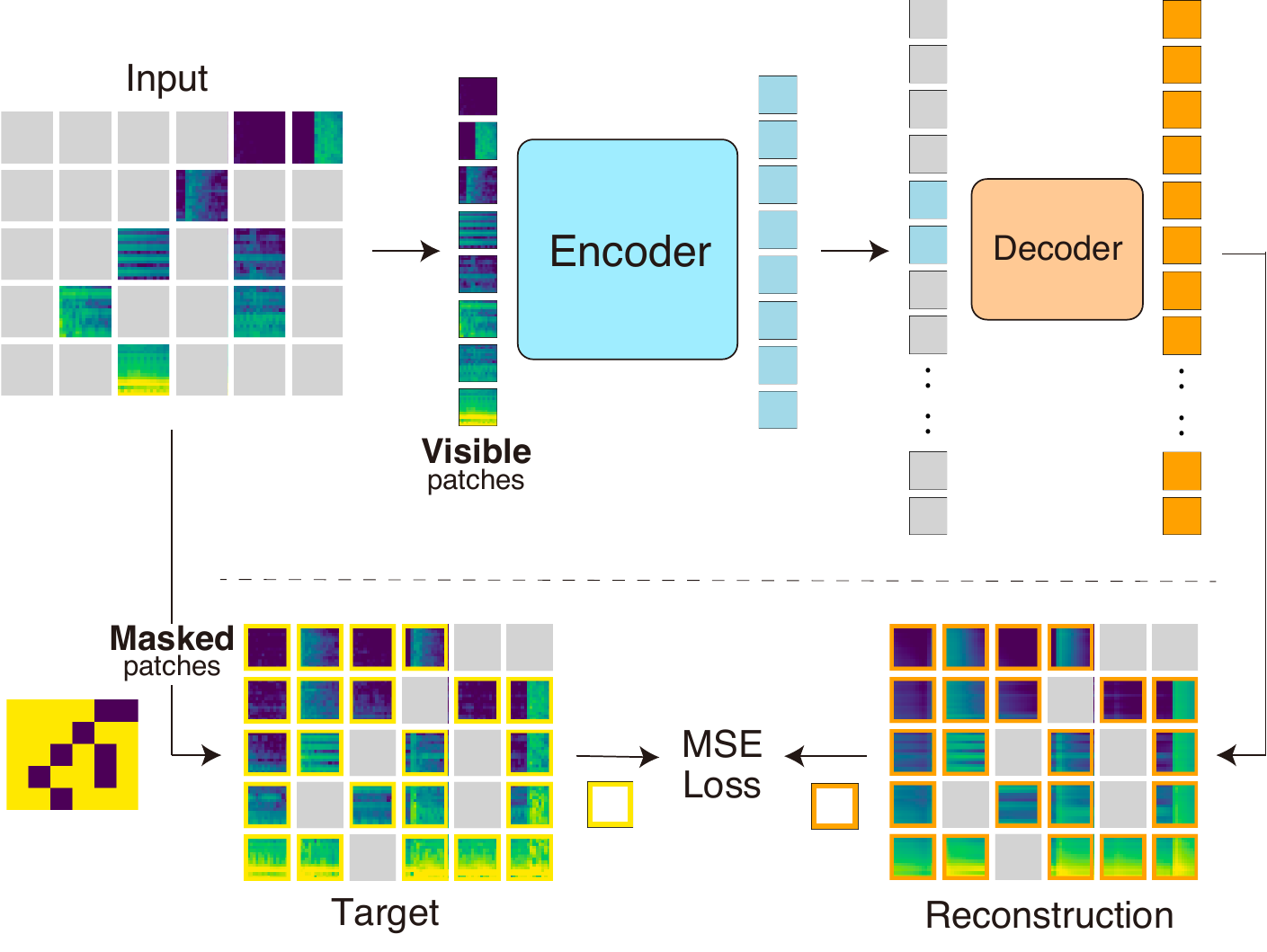}
  \caption{MAE \citep{he2021masked} pre-training flow; we redraw Figure 1 in the MAE paper, in which we replaced the input image with a spectrogram and added loss calculation flow. MAE masks 75\% of input patches, then the encoder processes the visible 25\% of patches only, saving 75\% of computation load. The lightweight decoder takes as input the encoded 25\% plus mask tokens that fill the masked 75\% of the input, then reconstructs spectrogram. The loss calculates the reconstruction error of the masked patches, which is a mean squared error (MSE).
  % TODO LOSS is using input as training target, the encoder is only used after training.
  }
  \label{fig:scenario}
  %\vspace{-10pt}
\end{figure}

These methods utilize self-supervising signals such as the temporal relationship between audio samples or differences in multiple audio samples generated by data augmentations.
For example, triplet loss or contrastive learning methods \citep{shor2020trill,saeed2020cola,spijkervet2021contrastive,fonseca2020uclser20} learn to make representations closer to temporally close audio segments while pushing away remote segments. Our previous study BYOL-A \citep{niizumi2021byol-a} learns representations invariant against the difference in audio signals created by data augmentations.

However, training signals of these methods do not provide information about complete details of the input.
The temporal relationship among audios can be the information about the difference between the audio signals that do not describe input details, and the data augmentations can change the details in original input audio signals.
Therefore, we think these training signals are suboptimal for learning to represent audio input as it is.

We think that the input signal itself can be the best training signal to learn representations that describe the input in detail.
Learning frameworks to achieve our goal include Masked Language Modeling (MLM) in natural language processing (NLP) or Masked Image Modeling (MIM) in the image domain.
These methods learn representations by masking a part of the input signal and using other parts to predict the masked signals.
In particular, MLM such as BERT \citep{bert} have already proven highly effective and have demonstrated strong performance.
Inspired by BERT, speech self-supervised learning methods that learn from masked input prediction \citep{baevski2020wav2vec2,Liu2020Mockingjay,Chi2021AudioAlbert,Hsu2021HuBERT}, have also shown solid results in the audio domain.
In the image domain, recent progress of MIM such as BEiT \citep{bao2021beit} and Masked Autoencoders (MAE) \citep{he2021masked} have shown promising performance such that \textit{“Self-supervised learning in vision may now be embarking on a similar trajectory as in NLP”} \citep{he2021masked}.
% liu2022audiosslsurvey

In this study, we explore the learning of general-purpose audio representations through the MIM applied to the audio spectrogram, which we call Masked Spectrogram Modeling (MSM).
MIM splits the input image into grid patches. Therefore, applying MIM to the audio spectrogram splits the input along the time and frequency axes, allowing \textit{“the model to learn both the temporal and frequency structure”} \citep{gong2021ssast}, unlike previous methods for speech (e.g., Mockingjay \citep{Liu2020Mockingjay} and wav2vec 2.0 \citep{baevski2020wav2vec2}) that split audio along time only.

To implement MSM, we use MAE as a training framework.
MAE learns to efficiently encode the small number of visible patches into latent representations to carry essential information for reconstructing masked patches, a large portion of the input signal.
It then calculates the reconstruction error as a training loss, achieving our goal of using the input itself as a training signal.

Our main contributions are the proposal and implementation of MSM using MAE (MSM-MAE) to learn general-purpose audio representations and results showing that our variants of MAE outperform other methods on some tasks in the HEAR 2021 NeurIPS Challenge \citep{turian2022hear}.
In addition, we investigate how the design choices of MSM-MAE impact the performance and present a qualitative analyses of learned representations using visualizations.
Our code is available online.

\section{Related Work}\label{sec:related}
\noindent\textbf{Audio representation learning closely related to our work.}
SSAST \citep{gong2021ssast} is a self-supervised learning method that pre-trains ViT \citep{ViT} using a pretext task of \textit{Joint Discriminative and Generative Masked Spectrogram Patch Modeling (MSPM)}, which combines contrastive learning and masked patch reconstruction. While the patch reconstruction task is the same as that with MAE, it uses a two-layer MLP to reconstruct masked patches, unlike MAE, which uses a sufficiently deep transformer.

Mockingjay \citep{Liu2020Mockingjay} proposed \textit{Masked Acoustic Model (MAM)}, a pretext task of reconstructing the masked time frames; the subsequent studies TERA \citep{Liu2021Tera} and Audio ALBERT \citep{Chi2021AudioAlbert} also use MAM.
Unlike MIM, MAM slices the spectrogram along time as natural handling of time-series data, samely as the methods that accept raw audio as input, such as wav2vec 2.0 \citep{baevski2020wav2vec2} and HuBERT \citep{Hsu2021HuBERT}.

PaSST \citep{koutini2021passt} is a supervised learning method that pre-trains ViT with the proposed \textit{Patchout}, taking a similar approach to MAE. The \textit{Patchout} reduces the number of patches to encode, thus, saving computation resources.

%For automatic speech recognition, wav2vec 2.0 \citep{baevski2020wav2vec2} and HuBERT learn contextualized speech representation using contrastive loss with masked tokens over a quantization of latent representations. The representations it learns are more discrete than describing the inputs as they are.

\noindent\textbf{Audio self-supervised learning.}
Methods based on triplet \citep{shor2020trill} or contrastive loss \citep{saeed2020cola, fonseca2020uclser20, spijkervet2021contrastive, wang2021universal} learn from the temporal relationship between audios; they learn to make feature embeddings of audios closer for the audios temporally closer, or push them away for the ones temporally remote.
Then, these methods do not model the representation of the input audio as it is.

On the other hand, our previous study BYOL-A \citep{niizumi2021byol-a} and SERAB BYOL-S \citep{scheidwasser2021serab} use data augmentations to produce multiple audios with a small degree of difference from the same audio input; they learn to make invariant representations for these augmented audios.
Therefore, these methods may not learn part of the audio in which data augmentations make changes.

\noindent\textbf{Masked Image Modeling.}
ViT \citep{ViT} conducts a self-supervised learning of predicting average 3bit color of masked patch as a preliminary exploration, resulted in 4\% behind supervised pre-training version of the ViT.
%  TODO describe the impact of 4% more -- with absolute 
BEiT \citep{bao2021beit} proposes a masked image modeling task that learns to predict discrete visual tokens of masked patches. BEiT outperforms supervised pre-trained ViT; however, it pre-trains the model to encode the input image into discrete tokens rather than representing the input as it is, and requires pre-trained discrete VAE \citep{ramesh2021dalle}, which is not available for audio.
%  (a discrete representation of patch pixels) 

\subsection{Masked Autoencoders (MAE)}\label{sec:mae}
MAE \citep{he2021masked} reconstructs the original signal given its partial observation.
Figure \ref{fig:scenario} illustrates the pre-training flow with a spectrogram as input.
First, MAE splits the input into patches and masks a large part of patches randomly, and then an encoder processes the visible patches only to latent representations.
Next, a decoder reconstructs input from the latent representations of visible patches and mask tokens representing masked patches.
Then the loss is calculated as the mean squared error (MSE) for all the masked patches between the reconstruction and target, which is normalized input.
After pre-training, only the encoder is applied, and it encodes whole patches of input images to produce representations for downstream tasks.

% an asymmetric design that allows the encoder to operate only on the partial, observed signal (without mask tokens) and a lightweight decoder that reconstructs the full signal from the latent representation and mask tokens. 

MAE masks a very large portion (e.g., 75\%) of patches with a notion that information density is different from that in languages and images; images are natural signals with spatial redundancy compared to languages, which are highly semantic and information-dense.
The MAE paper shows that the optimal mask ratio is 75\%, much higher than the 15\% of BERT \citep{bert} in the NLP domain.

Unlike classical autoencoders, MAE has an asymmetric encoder-decoder design; an encoder operates on the partial visible signal only, whereas a lightweight decoder reconstructs the full signal.
These design choices save computation load and enable us to scale MAE to train large models efficiently.

A sufficiently deep decoder is essential for linear evaluation performance without fine-tuning.
The last several layers in an autoencoder can be more specialized for reconstruction, thus becoming less relevant for other tasks.
A reasonably deep decoder can help make latent representations from the encoder output more abstract \citep{he2021masked,cao2022understand}.

%\subsection{Application to Spectrogram Audio}\label{sec:mae-spec}
\section{Masked Spectrogram Modeling using Masked Autoencoders}\label{sec:mae-spec}
We apply Masked Image Modeling to the audio spectrogram, which we call Masked Spectrogram Modeling (MSM).
MSM splits the input along the time and frequency axes, allowing it to learn both the temporal and frequency structure, unlike the previous methods for speech that split along time only.
To implement MSM, we use Masked Autoencoders (MSM-MAE).
In preliminary experiments, we found that MSM-MAE can follow the basic design choices of the original MAE, except the half decoder depth.
While the same mask ratio of 75\% suggests that information density is close to that of the image, the half decoder depth might suggest lower complexity of the context of the spectrogram than that of the image.

Besides the original designs, taking spectrogram as input introduces new choices, namely, input and patch size, because spectrogram has different axes of frequency and time, unlike an image.
For the input size, which consists of the number of frequency bins and time frames (denoted $F$ and $T$), we handle a constant $F$ and focus on exploring the optimal $T$.
%because the duration of the audio samples is variable in the downstream tasks.
We confirmed that both $T$ as input audio duration and patch size choices positively influence downstream task performance in preliminary experiments, unlike many other parameters that degrade performance with change.
Therefore, we investigate these two design choices in this paper.

In addition, to better use learned features in the downstream tasks, we also introduce a feature calculation specialized to the spectrogram input.

\subsection{Input Audio Duration}\label{sec:mae-input-duration}
During the pre-training, the longer duration, the model gains more chance to learn the relationships among the contents. Therefore, the longer duration could result in better representations. Meanwhile, the duration of samples in the downstream tasks ranges from 1-s to tens of seconds, or even longer. In addition, it may be fixed or vary from sample to sample. The optimal duration can depend on the task.

From the perspective of complexity, the shorter duration reduces the computation load of the two transformers on MAE because the length of the input sequence requires quadratic computational and memory complexity on the transformers; thus, the shorter duration is beneficial for scaling the system.
For these reasons, we study various input audio durations.

\subsection{Patch Size}\label{sec:mae-patch-size}

While both the input and patch sizes are square with the image, we handle rectangle input with the spectrogram. Thus, the same goes for the patch sizes.

The patch size also affects task performance because it sets the frequency/time resolution of encoded representations. There are various task demands; for example, pitch detection requires sufficient frequency resolution, whereas short event detection requires fine time resolution.
To meet these demands, we can make the patch smaller to make the resolution finer.

However, the available choice of resolution is limited due to computational complexity; for example, making the patch size half on both frequency and time will quadruple the sequence length.
We explore various patch sizes based on the default resolution of $16\times 16$.

\subsection{Feature Calculation for Downstream Tasks}\label{sec:msm-mae-feature-calc}
The encoder of the learned MAE encodes whole patches of the audio samples in the task, yielding embeddings of all patches for each audio sample; thereby, the embeddings for a single time frame consist of that of multiple patches of frequencies.

While typically, the patch embeddings for a time frame can be averaged to get a single embedding, we think it impairs available information by averaging embeddings among frequency bins.
Therefore, we calculate features by concatenating all the patch embeddings of the same time frame, preserving all available features as the following python pseudo code:

\begin{equation}
\begin{split}
z' = & z.\text{reshape}(B,  N_F, N_T, D)\\
    & .\text{transpose}(1, 2)\\
    &.\text{reshape}(B, N_T, N_F D)
\end{split}    
\end{equation}
where $z \in R^{B \times N_F N_T \times D}$ is the encoder output, $B$ is batch size, $N_F$ is the number of patches along frequency, $N_T$ is the number of patches along time, $D$ is a feature dimension, and $z' \in R^{B \times N_T \times N_F D}$ is the calculation result.
This calculation summarizes encoded features of a time frame for all frequencies into a single vector. For example, the feature dimension of $z'$ will be $3840$ when $D=768$ and $N_F=5$, which is used in our experiments.

\section{Experiments}\label{sec:experiments}
We evaluate our MSM-MAE models on a benchmark suite from the HEAR (Holistic Evaluation of Audio Representations) 2021 NeurIPS Challenge \citep{turian2022hear}, which spans multiple audio domains, including speech, environmental sound, and music.

We describe the details of experiments in Section \ref{sec:exp-details}, and the downstream tasks in \sectionref{sec:ds-tasks}.
Next, we evaluate our models with results on the HEAR 2021 Challenge in \sectionref{sec:exp-results-hear2021},
and investigate how design choices impact the performance in \sectionref{sec:exp-design-choice-impact}.
Then, we analyze learned representations qualitatively using visualizations in \sectionref{sec:qualitative-analysis}.

\subsection{Experimental Details}\label{sec:exp-details}
\subsubsection{Pre-training}
We used ViT-base \citep{ViT} as an encoder model and a smaller decoder with a width of 384-d, depth of 4, and 6 heads. Then, we pre-trained on MAEs with the original parameters except for the pre-training epoch of 100, warmup epoch of 10, batch size of 512, and learning rate of 6e-4.
While we normalized batch inputs with dataset statistics, we did not normalize the target when calculating reconstruction loss due to early observation of better performances.
We followed the original mask ratio of 0.75.

The pre-training dataset consisted of $1,963,807$ samples from balanced\_train\_segments and unbalanced\_train\_segments data splits of the AudioSet \citep{gemmeke2017audioset}. 
We preprocessed samples to a log-scaled mel spectrogram with a sampling frequency of 16,000 Hz, window size of 25 ms, hop size of 10 ms, and mel-spaced frequency bins $F=80$ in the range 50–8,000 Hz.

\subsubsection{Evaluation}
We evaluated models with input audio duration $T \in \{96,208,304,400,512\}$, corresponding to 960 ms to 5.12 seconds.
We also evaluated models with patch sizes ($F\times T$) of $16\times 16$ by default, $16\times 8$ or $16\times 4$ for double or quadruple time resolutions, and $8\times 16$ for a double frequency resolution.
In addition, we evaluated a model with a patch size of $80 \times 4$, which cuts the input along time, making spectrogram strips; note that we use the fixed number of frequency bins $F=80$.
The model configurations are listed in \tableref{tab:model-conf}.

\begin{table}[htbp]
\floatconts{tab:model-conf}{\caption{Model configuration details.}}{
\vspace{-0.2cm}
\resizebox{0.6\linewidth}{!}{%
\begin{tabular}{lrcccc}
\toprule
& \multicolumn{3}{c}{\# of Patches} & \multicolumn{2}{c}{Patch Size} \\
\cmidrule(lr){2-4} \cmidrule(lr){5-6}
Model &    Total &  Freq. &  Time &  Freq. &  Time \\
\midrule
MSM-MAE-96        &   30 &    5 &    6 &   16 &   16 \\
MSM-MAE-208       &   65 &    5 &   13 &   16 &   16 \\
MSM-MAE-304       &   95 &    5 &   19 &   16 &   16 \\
MSM-MAE-400       &  125 &    5 &   25 &   16 &   16 \\
MSM-MAE-512       &  160 &    5 &   32 &   16 &   16 \\
MSM-MAE-200 ($16\times 8$) &  125 &    5 &   25 &   16 &    8 \\
MSM-MAE-200 ($16\times 4$) &  250 &    5 &   50 &   16 &    4 \\
MSM-MAE-208 ($8\times 16$) &  130 &   10 &   13 &    8 &   16 \\
MSM-MAE-304 ($80 \times 4$) & 76 & 1 & 76 & 80 & 4 \\
\bottomrule
\end{tabular}
}}
\end{table}

We used the \textit{hear-eval-kit}\footnote{\url{https://github.com/neuralaudio/hear-eval-kit}} from the HEAR 2021 Challenge.
The \textit{hear-eval-kit} evaluates the performance of the models on the downstream tasks without fine-tuning.
First, it encodes all the task samples into embeddings using the model as a feature extractor. Then, it trains a shallow downstream model to solve the task, taking the embeddings as input. It reports the test performance of the downstream model as the model performance.

The \textit{hear-eval-kit} requires two types of embeddings from models. One is \textit{timestamp embeddings}, for which we used the $z'$, features calculated for the downstream task described in \sectionref{sec:msm-mae-feature-calc}; the other is \textit{scene embeddings}, for which we calculated the temporal average of $z'$.
Since the ViT model accepts the fixed input duration $T$, we convert the variable-length inputs into feature vectors in two steps: encode all the divided segments of length $T$ of input, and then concatenate the encoded features along time.
All other details of downstream task evaluation follow the defaults of the \textit{hear-eval-kit}, including the network design of the shallow downstream models for each task.

\subsection{Downstream Tasks}\label{sec:ds-tasks}
We used 15 downstream tasks from the HEAR 2021 \citep{turian2022hear}, consisting of four environmental sound tasks, five speech tasks, and six music tasks.

\noindent\textbf{Speech tasks.}
These tasks are for non-semantic speeches, which do not include automatic speech recognition:
Speech Commands \citep{speechcommandsv2} (SPC, speech command word classification),
CREMA-D \citep{cao2014cremad} (CRM-D, speech emotion recognition),
LibriCount \citep{libricount} (LbCount, speaker count estimation),
Vocal Imitations \citep{VocalImitation} (VoImit, matching a vocal imitation with an original sound as a classification),
and Vox Lingua Top 10 \citep{VoxLingua107} (Lingua10, language identification).

\noindent\textbf{Environmental sound tasks.}
ESC-50 \citep{piczak2015esc50} (environmental sound classification),
FSD50K \citep{fonseca2020fsd50k} (multilabel sound event classification),
Gunshot Triangulation \citep{cooper2020gunshots} (Gunshot, recording location classification),
Beehive States \citep{beehivestates} (Beehive, normal or queen-less binary classification).

\noindent\textbf{Music tasks.}
GTZAN \citep{gt2013gtzan} (music genre recognition),
GTZAN Music Speech (GTZ-M/S, music or speech binary classification),
NSynth \citep{nsynth2017} Pitch (NSPitch, pitch classification),
Mridingham Stroke and Tonic \citep{mridangam} (Mrd-Stk for stroke, or Mrd-Ton for tonic, pitched percussion stroke or tonic classification),
and Beijing Opera Percussion \citep{beijingopera} (Beijing, percussion instrument classification).

\subsection{Experimental Results: Comparison with the HEAR 2021 Results}\label{sec:exp-results-hear2021}
We compare the results of our two best performing models with the results from the HEAR 2021 Challenge in \tableref{tab:result-esc-hear,tab:result-spc-hear,tab:result-music-hear}.
We used the HEAR 2021 Challenge results for which a single model is used---an ensemble of models is out of the scope of this study---and whose details are available in their papers.

\begin{table*}[htbp]
\floatconts{tab:result-esc-hear}{\caption{Environmental sound task results.}}{
\vspace{-0.2cm}
\resizebox{0.85\linewidth}{!}{%
\begin{tabular}{lllll}
\toprule
Model &     Gunshot &      FSD50K &      ESC-50 &     Beehive \\
\midrule
CREPE \citep{kim2018crepe}                 &        86.3 &        15.9 &        30.0 &        59.3 \\
wav2vec2 \citep{baevski2020wav2vec2}       &        84.8 &        11.6 &        56.1 &         N/A \\
KW-MLP \citep{morshed2021kwmlp}            &        93.2 &        18.7 &        36.7 &        76.0 \\
SERAB BYOL-S \citep{scheidwasser2021serab} &        85.7 &        50.9 &        80.5 &        54.9 \\
OpenL3 \citep{cramer2019openl3}            &        94.9 &        44.7 &        75.1 &        60.4 \\
Wav2CLIP \citep{wu2021wav2clip}            &        92.9 &        36.2 &        75.9 &  \textbf{77.0} \\
PANNs CNN14 \citep{kong2020panns}          &        79.8 &         N/A &        90.9 &        44.6 \\
PaSST base \citep{koutini2021passt}        &        94.0 &  \textbf{64.0} &  \textbf{94.7} &         N/A \\
\midrule
MSM-MAE-200 ($16\times 4$) (ours) &        95.2 &        50.9 &        84.3 &        60.4 \\
MSM-MAE-512 (ours)        &  \textbf{96.4} &        52.2 &        85.6 &        69.4 \\
\bottomrule
\end{tabular}
}}
\end{table*}

\begin{table*}[htbp]
\vspace{-0.2cm}
\floatconts{tab:result-spc-hear}{\caption{Speech task results.}}{
\vspace{-0.2cm}
\resizebox{0.85\linewidth}{!}{%
\begin{tabular}{llllll}
\toprule
Model &     Lingua10 &      VoImit &       CRM-D &         SPC &     LbCount \\
\midrule
CREPE \citep{kim2018crepe}                 &        14.2 &         5.1 &        38.3 &        21.1 &        49.9 \\
Wav2CLIP \citep{wu2021wav2clip}            &        19.2 &         8.3 &        51.2 &        34.7 &        52.8 \\
KW-MLP \citep{morshed2021kwmlp}            &        18.1 &         5.6 &        42.4 &  \textbf{97.8} &        45.1 \\
PANNs CNN14 \citep{kong2020panns}          &        24.4 &        12.7 &        55.5 &        61.8 &        65.2 \\
PaSST base \citep{koutini2021passt}        &        25.9 &        18.2 &        61.0 &        63.9 &        66.0 \\
OpenL3 \citep{cramer2019openl3}            &        33.1 &         7.8 &        55.0 &        76.3 &        64.1 \\
wav2vec2 \citep{baevski2020wav2vec2}       &        49.3 &         8.0 &        65.6 &        87.9 &        69.2 \\
SERAB BYOL-S \citep{scheidwasser2021serab} &        45.8 &        16.0 &        65.7 &        94.8 &        78.5 \\
\midrule
MSM-MAE-200 ($16\times 4$) (ours)                   &  \textbf{52.9} &        12.8 &        73.3 &        87.3 &  \textbf{85.8} \\
MSM-MAE-512 (ours)                          &        50.0 &  \textbf{18.3} &  \textbf{73.4} &        86.4 &        77.8 \\
\bottomrule
\end{tabular}
}}
\end{table*}

\begin{table*}[htbp]
\vspace{-0.1cm}
\floatconts{tab:result-music-hear}{\caption{Music task results.}}{
\vspace{-0.2cm}
\resizebox{0.97\linewidth}{!}{%
\begin{tabular}{lllllllll}
\toprule
Model &      GTZ-M/S &       GTZAN &     NSPitch &     Mrd-Ton &     Mrd-Stk &     Beijing \\
\midrule
PANNs CNN14 \citep{kong2020panns}          &  \textbf{99.2} &        86.0 &        30.1 &        82.4 &        93.9 &        91.1 \\
Wav2CLIP \citep{wu2021wav2clip}            &        94.6 &        74.8 &        43.9 &        82.9 &        94.7 &        93.6 \\
KW-MLP \citep{morshed2021kwmlp}            &        88.9 &        55.4 &        60.5 &        94.2 &        96.9 &        91.1 \\
wav2vec2 \citep{baevski2020wav2vec2}       &        94.6 &        78.0 &        65.3 &        82.8 &        94.3 &        90.7 \\
CREPE \citep{kim2018crepe}                 &        92.9 &        64.5 &  \textbf{90.0} &        82.4 &        89.8 &        92.8 \\
PaSST base \citep{koutini2021passt}        &        97.7 &  \textbf{88.3} &        54.1 &        81.9 &        96.5 &        96.6 \\
SERAB BYOL-S \citep{scheidwasser2021serab} &        93.8 &        83.7 &        71.2 &        92.8 &        97.3 &        95.3 \\
OpenL3 \citep{cramer2019openl3}            &        96.9 &        87.9 &        73.1 &        93.7 &        96.7 &  \textbf{97.5} \\
\midrule
MSM-MAE-200 ($16\times 4$) (ours)                  &        97.7 &        86.3 &        84.0 &  \textbf{98.6} &  \textbf{97.5} &        95.3 \\
MSM-MAE-512 (ours)                          &        99.2 &        86.1 &        81.2 &        98.3 &        97.5 &        94.9 \\
\bottomrule
\end{tabular}
}}
\end{table*}

These tables show that our models outperform others in seven tasks: Gunshot Triangulation, Vox Lingua Top 10, Vocal Imitation, CREMA-D, LibriCount, and Mridingham Tonic and Stroke.
Conversely, models specialized in the task outperform our models on other tasks. On environmental sound tasks, FSD50K and ESC-50, the PANNs and PaSST-base outperform ours. These models are supervised learning pre-trained on the AudioSet using its labels.
KW-MLP, wav2vec2, and SERAB-BYOLS outperform ours on the Speech Commands task. These models are pre-trained on a speech corpus for specialization in speech.
CREPE, specializing in pitch estimation, outperforms ours on the NSynth Pitch task.
With the exception of these specialized models, our MSM-MAE models show the top results in most tasks.
These results show that MSM-MAE learns audio representation effective for general tasks without specializing in domains.

\subsection{Experimental Results: Impact of Design Choices on Performance}\label{sec:exp-design-choice-impact}
We explore the impact of the design choices of input audio duration and patch size, for which we found positive effects in preliminary experiments.
In addition, we also evaluate the performance difference between designs for splitting the input.

\subsubsection{Impact of Input Audio Duration}
\tableref{tab:result-duration} shows the results of MSM-MAE with various input audio durations.
The number at the tail of the model name shows the duration, which is $10\times$ in ms in actual time (e.g., 96 and 512 are 960 ms and 5.12 seconds, respectively).

These results show that the longer durations yield better results among HEAR 2021 tasks. We see an exceptional behavior with speech tasks (Lingua10, SPC, and LbCount), Gunshot, and GTZ-M/S. However, we think that a long input duration is more beneficial in general-purpose use, because most tasks show the best result with the longer input duration of 400 or 512.

\begin{table*}[htbp]
\vspace{-0.2cm}
\floatconts{tab:result-duration}{\caption{Task results for various input audio durations.}}{
\vspace{-0.2cm}
\subtable[Environmental sound tasks]{
\resizebox{0.5\linewidth}{!}{%
\begin{tabular}{lllll}
\toprule
Model &     Gunshot &      FSD50K &      ESC-50 &     Beehive \\
\midrule
MSM-MAE-96  &  \textbf{98.8} &        48.7 &        79.9 &        54.8 \\
MSM-MAE-208 &        89.9 &        50.8 &        84.9 &        52.0 \\
MSM-MAE-304 &        90.5 &        51.7 &        85.3 &        65.3 \\
MSM-MAE-400 &        90.5 &        51.8 &  \textbf{85.6} &        62.6 \\
MSM-MAE-512 &        96.4 &  \textbf{52.2} &        85.6 &  \textbf{69.4} \\
\bottomrule
\end{tabular}
}}
\subtable[Speech tasks]{
\resizebox{0.6\linewidth}{!}{%
\begin{tabular}{llllll}
\toprule
Model &     Lingua10 &      VoImit &       CRM-D &         SPC &     LbCount \\
\midrule
MSM-MAE-96  &        39.0 &        12.6 &        68.0 &        86.6 &        78.7 \\
MSM-MAE-208 &        48.2 &        15.7 &        70.4 &  \textbf{87.5} &        80.3 \\
MSM-MAE-304 &  \textbf{50.6} &        17.2 &        72.0 &        87.3 &  \textbf{80.6} \\
MSM-MAE-400 &        45.9 &        16.9 &        72.8 &        85.8 &        80.1 \\
MSM-MAE-512 &        50.0 &  \textbf{18.3} &  \textbf{73.4} &        86.4 &        77.8 \\
\bottomrule
\end{tabular}
}}
\subtable[Music tasks]{
\resizebox{0.7\linewidth}{!}{%
\begin{tabular}{lllllllll}
\toprule
Model &      GTZ-M/S &       GTZAN &     NSPitch &     Mrd-Ton &     Mrd-Stk &     Beijing \\
\midrule
MSM-MAE-96  &         96.9 &        84.4 &        81.0 &        98.2 &        97.3 &        89.8 \\
MSM-MAE-208 &  \textbf{100.0} &        84.9 &        81.3 &        98.3 &        97.5 &        92.8 \\
MSM-MAE-304 &         98.4 &        85.6 &        81.2 &        98.1 &        97.5 &        94.9 \\
MSM-MAE-400 &         98.5 &        86.0 &  \textbf{81.6} &  \textbf{98.5} &  \textbf{97.7} &        94.5 \\
MSM-MAE-512 &         99.2 &  \textbf{86.1} &        81.2 &        98.3 &        97.5 &  \textbf{94.9} \\
\bottomrule
\end{tabular}
}}
}
\end{table*}

\subsubsection{Impact of Patch Size}
\tableref{tab:result-patchsize} shows the results of MSM-MAE with various patch sizes for the models that accept 2-s input audio durations: MSM-MAE-208 ($16\times 16$), MSM-MAE-200 ($16\times 8$), MSM-MAE-200 ($16\times 4$), and MSM-MAE-208 ($8\times 16$).
The $16\times 16$, $16\times 8$, $16\times 4$, and $8\times 16$ show the patch size of $F$ by $T$.

\begin{table*}[htbp]
\vspace{-0.2cm}
\floatconts{tab:result-patchsize}{\caption{Task results for various patch sizes. $N$ is the total number of patches.}}{
\vspace{-0.2cm}
\subtable[Environmental sound tasks]{
\resizebox{0.7\linewidth}{!}{%
\begin{tabular}{lllll}
\toprule
Model &     Gunshot &      FSD50K &      ESC-50 &     Beehive \\
\midrule
MSM-MAE-208 ($16\times 16, N=65$) &        89.9 &        50.8 &  \textbf{84.9} &        52.0 \\
MSM-MAE-200 ($16\times 8, N=125$)  &        94.0 &        51.1 &        83.9 &        62.0 \\
MSM-MAE-200 ($16\times 4, N=250$)  &        95.2 &        50.9 &        84.3 &        60.4 \\
MSM-MAE-208 ($8\times 16, N=130$)  &  \textbf{96.4} &  \textbf{51.2} &        84.6 &  \textbf{62.0} \\
\bottomrule
\end{tabular}
}}
\subtable[Speech tasks]{
\resizebox{0.8\linewidth}{!}{%
\begin{tabular}{llllll}
\toprule
Model &     Lingua10 &      VoImit &       CRM-D &         SPC &     LbCount \\
\midrule
MSM-MAE-208 ($16\times 16, N=65$) &        48.2 &  \textbf{15.7} &        70.4 &        87.5 &        80.3 \\
MSM-MAE-200 ($16\times 8, N=125$)  &        52.4 &        14.7 &        73.1 &  \textbf{87.7} &        84.5 \\
MSM-MAE-200 ($16\times 4, N=250$)  &  \textbf{52.9} &        12.8 &  \textbf{73.3} &        87.3 &  \textbf{85.8} \\
MSM-MAE-208 ($8\times 16, N=130$)  &        48.4 &        14.1 &        71.4 &        87.0 &        82.8 \\
\bottomrule
\end{tabular}
}}
\subtable[Music tasks]{
\resizebox{0.93\linewidth}{!}{%
\begin{tabular}{lllllllll}
\toprule
Model &      GTZ-M/S &       GTZAN &     NSPitch &     Mrd-Ton &     Mrd-Stk &     Beijing \\
\midrule
MSM-MAE-208 ($16\times 16, N=65$) &  \textbf{100.0} &        84.9 &        81.3 &        98.3 &        97.5 &        92.8 \\
MSM-MAE-200 ($16\times 8, N=125$)  &         99.2 &        85.7 &        81.7 &  \textbf{98.6} &  \textbf{97.6} &        93.6 \\
MSM-MAE-200 ($16\times 4, N=250$)  &         97.7 &        86.3 &  \textbf{84.0} &        98.6 &        97.5 &  \textbf{95.3} \\
MSM-MAE-208 ($8\times 16, N=130$)  &         98.5 &  \textbf{86.4} &        83.2 &        98.4 &        97.5 &        92.8 \\
\bottomrule
\end{tabular}
}}
}
\end{table*}

The results show that the finer time resolutions ($16\times 8$ and $16\times 4$) improve performance on 11 out of 15 tasks (Gunshot, FSD50K, Beehive, Lingua10, CRM-D, LbCount, GTZAN, NSPitch, Mrd-Ton, Mrd-Stk, and Beijing). In contrast, the finer frequency resolution ($8\times 16$) improves eight tasks (Gunshot, FSD50K, Beehive, CRM-D, LbCount, GTZAN, NSPitch, and Mrd-Ton), and the degree of improvements are smaller than for the finer time resolutions.

We can also find the different trends in improvements among models.
If we focus on the best results (bold numbers), the finer time resolutions ($16\times 8$ and $16\times 4$) excel on speech and music tasks, whereas the finer frequency resolution ($8\times 16$) performs better on environmental sound tasks.

In summary, we can improve task performance with finer resolutions using smaller patch sizes, especially with finer time resolutions.
However, $16\times 4$ shows that it is not always the case compared to $16\times 8$; the finer $16\times 4$ does not clearly show performance improvements superior to that for $16\times 8$, even though the $16\times 4$ costs quadratically more computational complexity than $16\times 8$.

\subsubsection{Impact of Input Splitting: Patches vs. Strips}
\tableref{tab:result-patch-vs-strip} shows the results of MSM-MAE for different designs of splitting the input spectrogram into patches or strips. MSM-MAE-304 ($16 \times 16$) cuts the input spectrogram into patches along both frequency and time.
In contrast, MSM-MAE-304 ($80 \times 4$) cuts the input into strips along time, simulating the previous methods such as Mockingjay \citep{Liu2020Mockingjay} that handle the input as a sequence of spectrogram strips. We compare the performance difference between these two ways.

\begin{table*}[htbp]
\vspace{-0.2cm}
\floatconts{tab:result-patch-vs-strip}{\caption{Task results for comparing patches vs. strips.}}{
\vspace{-0.2cm}
\subtable[Environmental sound tasks]{
\resizebox{0.65\linewidth}{!}{%
\begin{tabular}{lllll}
\toprule
Model &     Gunshot &      FSD50K &      ESC-50 &     Beehive \\
\midrule
MSM-MAE-304 ($16 \times 16$, Patches) &  \textbf{90.5} &  \textbf{51.7} &  \textbf{85.3} &        65.3 \\
MSM-MAE-304 ($80 \times 4$, Strips)  &        85.7 &        49.1 &        81.5 &  \textbf{66.0} \\
difference          &         4.8 &         2.6 &         3.8 &        -0.7 \\
\bottomrule
\end{tabular}
}}
\subtable[Speech tasks]{
\resizebox{0.75\linewidth}{!}{%
\begin{tabular}{llllll}
\toprule
Model &     Lingua10 &      VoImit &       CRM-D &         SPC &     LbCount \\
\midrule
MSM-MAE-304 ($16 \times 16$, Patches) &  \textbf{50.6} &  \textbf{17.2} &        72.0 &        87.3 &  \textbf{80.6} \\
MSM-MAE-304 ($80 \times 4$, Strips)  &        50.0 &        16.8 &  \textbf{73.8} &  \textbf{88.1} &        79.7 \\
difference          &         0.6 &         0.3 &        -1.9 &        -0.8 &         0.9 \\
\bottomrule
\end{tabular}
}}
\subtable[Music tasks]{
\resizebox{0.9\linewidth}{!}{%
\begin{tabular}{lllllllll}
\toprule
Model &      GTZ-M/S &       GTZAN &     NSPitch &     Mrd-Ton &     Mrd-Stk &     Beijing \\
\midrule
MSM-MAE-304 ($16 \times 16$, Patches) &  \textbf{98.4} &  \textbf{85.6} &  \textbf{81.2} &  \textbf{98.1} &        97.5 &  \textbf{94.9} \\
MSM-MAE-304 ($80 \times 4$, Strips)  &        96.9 &        85.5 &        81.0 &        98.1 &  \textbf{97.6} &        94.5 \\
difference          &         1.5 &         0.1 &         0.1 &         0.1 &        -0.1 &         0.4 \\
\bottomrule
\end{tabular}
}}
}
\end{table*}

The results show that the patch input model ($16 \times 16$) outperforms strip input model ($80 \times 4$) on most tasks.
It could indicate that the MIM framework, which takes patches as input and learns both frequency and time structure, is more suitable for learning general-purpose audio representations than the training framework that takes sequential strips as input.
By contrast, the strip input model outperforms the patch input model on CREMA-D and SPC tasks, which could also indicate that the strip input is effective for speech tasks, as used in the previous speech self-supervised learning studies that take sequential spectrogram strips as input.

\subsection{Qualitative Analysis with Visualizations}\label{sec:qualitative-analysis}
This section presents analyses based on visualizations to gain an understanding of the representations learned by MSM-MAE. \sectionref{sec:viz-recon-random,sec:viz-recon-pattern,sec:viz-recon-ratio} present visualization of reconstruction results, and \sectionref{sec:viz-att} shows visualizations of attention maps.
We present more visualizations in \appendixref{appendix:viz-recon}.

\begin{figure*}[htbp]
\floatconts{fig:viz-recon-rand3}
{\caption{Reconstructions for the pre-trained MSM-MAE-304 with a mask ratio of 0.75, obtained from three attempts of reconstruction of three sounds. Each example shows the input, the reconstruction result, and the difference between them, the error (RMS); the darker the color, the higher the reconstruction error.
The white squares in the reconstruction results show the visible patch.}}
{\includegraphics[width=\textwidth]{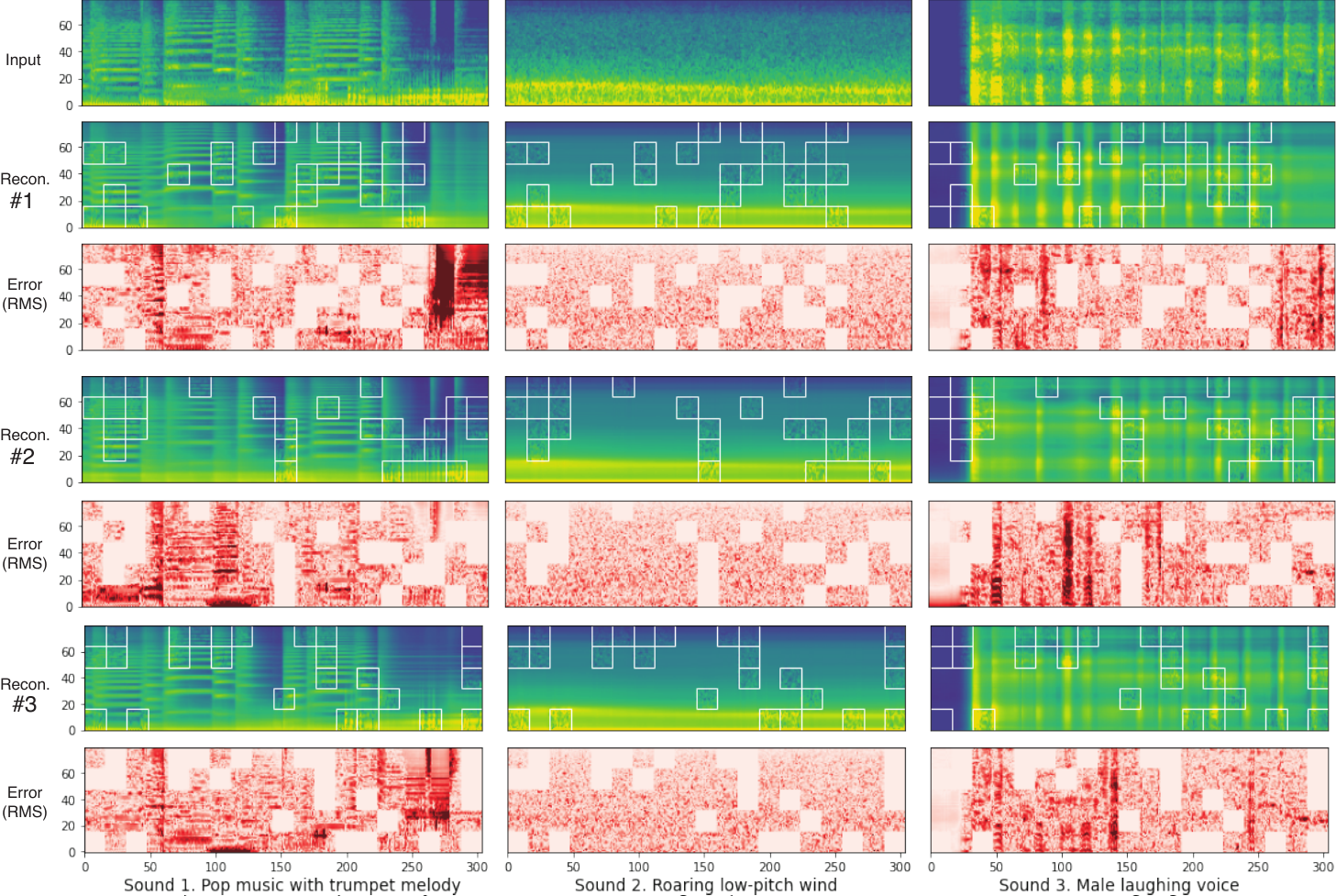}
\vspace{-30pt}
}
\vspace{-10pt}
\end{figure*}

\subsubsection{Reconstructions with Random Masks}\label{sec:viz-recon-random}
The reconstructions of three sounds in \figureref{fig:viz-recon-rand3} show results similar to those in the MAE paper \citep{he2021masked}, which reconstruct inputs well but with blurry details, indicating that our models were successfully trained in the experiments.
In \figureref{fig:viz-recon-rand3}, we observe that the frequency structures, especially the harmonic structures, are reconstructed, and the stationary sounds are easy to reconstruct compared to the short sound events.

Frequency structure reconstructions can be observed by focusing on reconstruction along the frequency axis (vertical axis).
Frequency bins are reconstructed even where a few visible patches (white squares) are available in a time frame.
Furthermore, we can find clear patterns of the harmonic structures in the reconstruction of sound 1 with trumpet notes.
These observations show that the latent representation is encoded effectively to reconstruct frequency bins in a time frame using limited information of a few patches.
%In contrast, we observe that sound events disappear where no visible patch is available in a frame on sounds 1 and 3, consisting of many sound events.
%Interestingly, we find clear patterns of the harmonic structures in the reconstruction of sound 1 with trumpet notes. These patterns suggest that the learned representation contains information useful to represent the pitch of the sound.

We find that stationary sound can be easily reconstructed from sound 2, which is the sound of roaring low-pitch wind.
The error of sound 2 shows a lighter color compared to the sound 1 or 3, indicating that the reconstructing of the 2 made a smaller error.
In addition, frequency structures are recovered entirely, even without an available visible patch in a time frame, contrary to the sound 1 and 3, where some sound events are not recovered.
These observations suggest that the learned representations carry the information related to temporal structure for each sound, such as stationary or short events.

\begin{figure*}[htbp]
\floatconts{fig:viz-recon-randhalf}
{\caption{Reconstructions with three patterns of masks, showing the difference in the error under different availability of visible information along axes.
The mask ratio is 0.5, i.e., half of the patches are masked for all examples.
%A, B, and C represent the Vertical mask with alternating masked and visible patches along the time axis, the Horizontal mask with alternating masked and visible patches along the frequency axis, and the Chessboard mask with masked and visible patches adjacent to both axes, respectively.
}}
{\includegraphics[width=\textwidth]{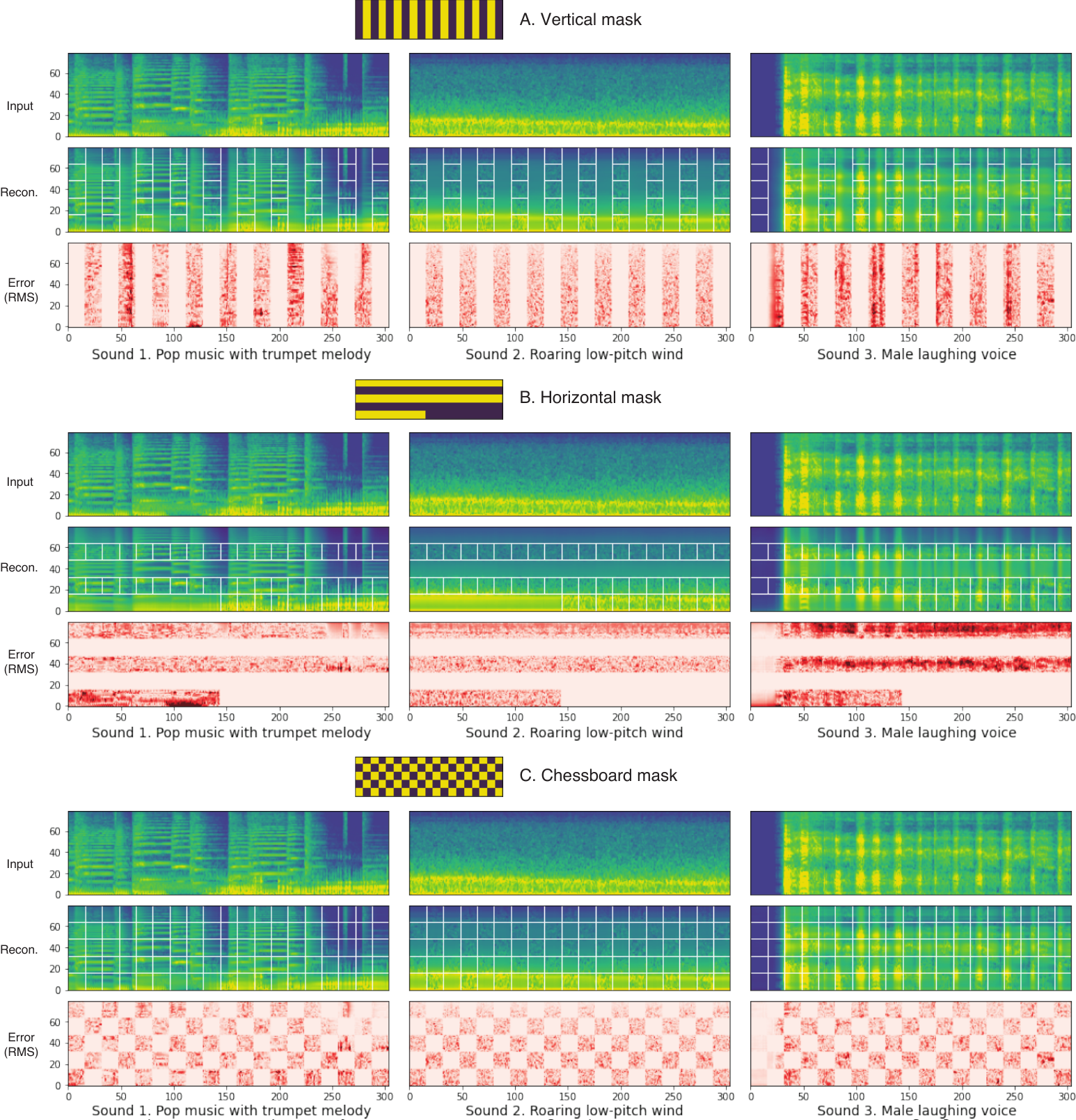}
\vspace{-30pt}
}
\vspace{-10pt}
\end{figure*}

\subsubsection{Reconstructions with Patterned Masks}\label{sec:viz-recon-pattern}
We compare the difference in reconstruction under the different availability of visible information made by the three mask patterns in \figureref{fig:viz-recon-randhalf}.
In A, vertical mask, visible and mask patches alternate along the time axis; models are to recover masked patches using visible patches adjoining on the time axis.
In B, horizontal mask, they alternate along the frequency axis; visible patches adjoining in the frequency axis are available.
In C, the chessboard mask is for making visible patches available around the masked patches.
All cases have the same mask ratio of 0.5.

In \figureref{fig:viz-recon-randhalf}, we observe that the more adjoining visible patches available, the easier the reconstruction becomes.
For all the sounds, the errors show lighter color on the chessboard mask results than that on the other vertical and horizontal mask results, showing the smaller reconstruction error for the chessboard mask, where more adjoining visible patches are available.

If we focus on the horizontal mask in B, we can observe that harmonic structures are reconstructed more easily than noises. Sound 1, a rich harmonic structure of trumpet notes, shows less reconstruction error than sound 3, with less structured frequency patterns of noises in laughing voices.
This observation suggests that the learned representation encodes the information of the harmonic structure more effectively than that of the noise.

\subsubsection{Reconstructions with Various Mask Ratios}\label{sec:viz-recon-ratio}
We varied the mask ratio to observe how the reconstruction changes according to it.
We used the 3-s model (MSM-MAE-304) to reconstruct the input spectrograms with varied mask ratios from 0.40 to 0.99, focusing on cases with extremely small numbers of visible patches (e.g., 1, 2, 5, and 10 visible patches, corresponding to mask ratios of 0.99, 0.98, 0.95, and 0.90, respectively).

\figureref{fig:viz-recon-varrate} shows the example reconstruction results.
The results show that reconstruction succeeds entirely up to the default ratio of 0.75, whereas it degrades noticeably at higher ratios.
As the mask ratio increases, only the patches around the visible patch are reconstructed relatively accurately, while the rest become blurry copies of the pattern around the nearest visible patch.

Focusing on the mask ratio of 0.99, which encodes only one visible patch, we can observe the reconstruction of both local and global patterns.
Locally, the frequency pattern of the time frames around the visible patch is reconstructed relatively clearly.
Globally, a frequency pattern similar to the average of the original spectrogram is reconstructed stationary over the entire spectrogram.
This observation suggests that even though only one patch was encoded, information related to both local and global patterns learned from the training dataset is embedded in the representation.

%The results with mask ratios of 0.99, where only one visible patch is available, show how the patch reconstruction process uses the information about the learned frequency and temporal patterns.
%The sound 1 result shows harmonic patterns around the visible patch along the frequency axis and stationary temporal patterns from the beginning until the tail of the spectrogram.
%The sound 3 result, a male laughing voice, shows the temporal repetition of a laugh voice pattern, and the sound 2 result is also similar.
%In these observations, the frequential patterns are prominent around the visible patch but blurry with more remote patches, whereas temporal patterns continue towards both ends on the time axis.
%Overall, it is clear that reconstruction of entire patches uses the information of the visible patches as well as the distances in the frequential and temporal positions.
%These observations may imply that the learned representation could hold information related to pitch and rhythm and may also explain the strong results of MSM-MAE on HEAR 2021 Challenge tasks in \sectionref{sec:exp-results-hear2021}, where this information is considered useful.

\begin{figure*}[htbp]
\floatconts{fig:viz-recon-varrate}
{\caption{Reconstructions with various mask ratios, focusing on cases of extremely small numbers of visible patches. We show white squares, the visible patches, on the results with mask ratios higher than 0.75 only to improve visibility. MSE is the mean squared error of reconstruction averaged over the entire spectrogram.}}
{\includegraphics[width=\textwidth]{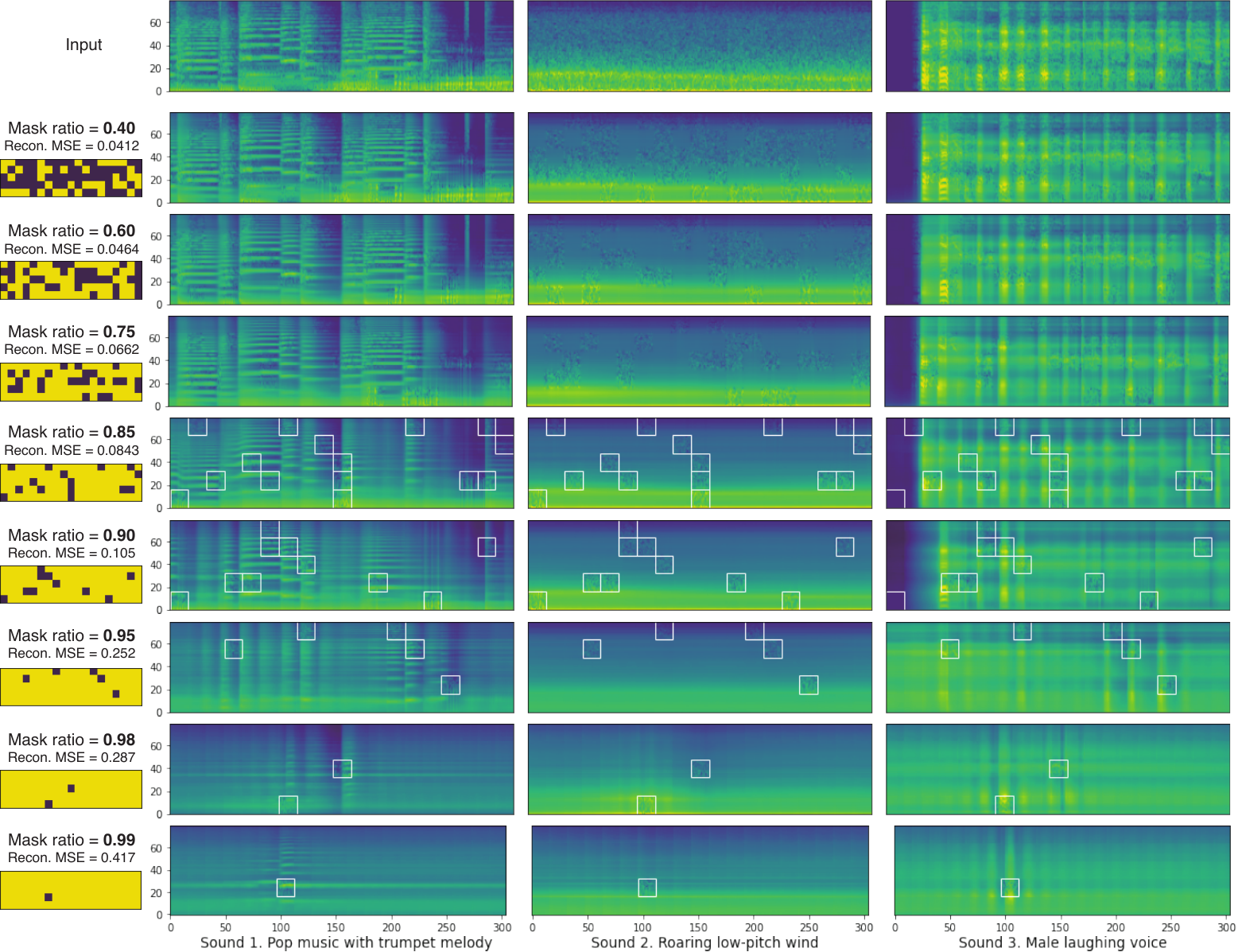}
\vspace{-30pt}
}
\vspace{-10pt}
\end{figure*}

\subsubsection{Self-Attention Map Visualizations}\label{sec:viz-att}
We visualize the self-attention of the encoder of pre-trained MSM-MAE-304 for six sounds in \figureref{fig:viz-attn-1,fig:viz-attn-2}.
We picked two reference points each to show self-attention maps.  These self-attention maps average from all heads in the last layer.

\begin{figure*}[htbp]
\floatconts{fig:viz-attn-1}
{\caption{MSM-MAE encoder self-attention map for reference points (2, 7) and (1, 12).}}
{\includegraphics[width=\textwidth]{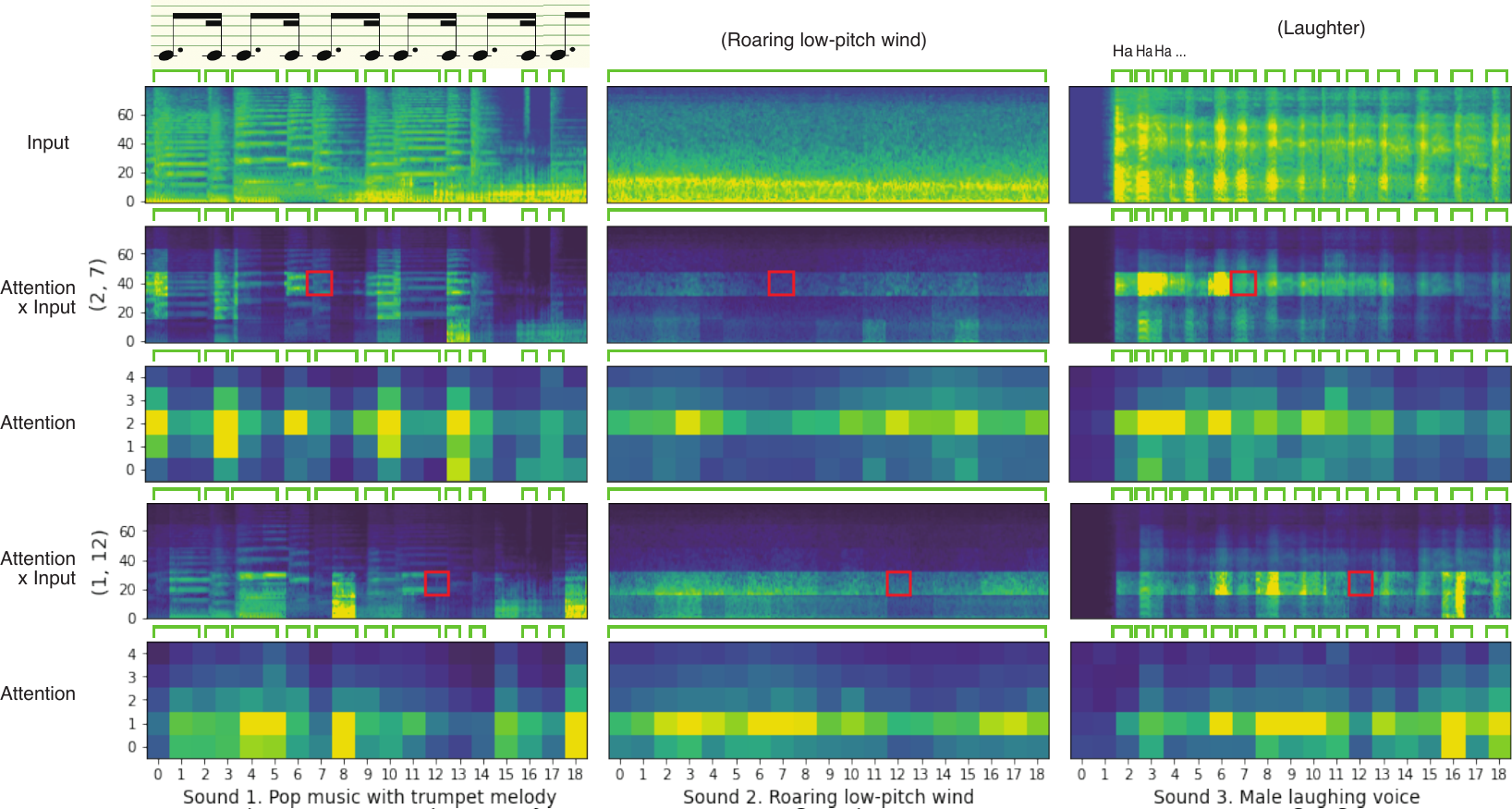}
\vspace{-30pt}
}
\vspace{-5pt}
\end{figure*}

We see that the self-attention map reflects the repetition or continuation in the input sounds.
Sound 1 of pop music has a clear repetition of notes in the input, which we can also find in the self-attention map.
On the other hand, sound 2 is a stationary sound of blowing wind, and the attention continues along time similarly.
Interestingly, we find in the first reference point in sound 1 that the self-attentions are strong at the same position in the beat, even though there are two similar notes per beat.

\begin{figure*}[htbp]
\floatconts{fig:viz-attn-2}
{\caption{MSM-MAE encoder self-attention map for reference points (0, 1) and (1, 7).}}
{\includegraphics[width=\textwidth]{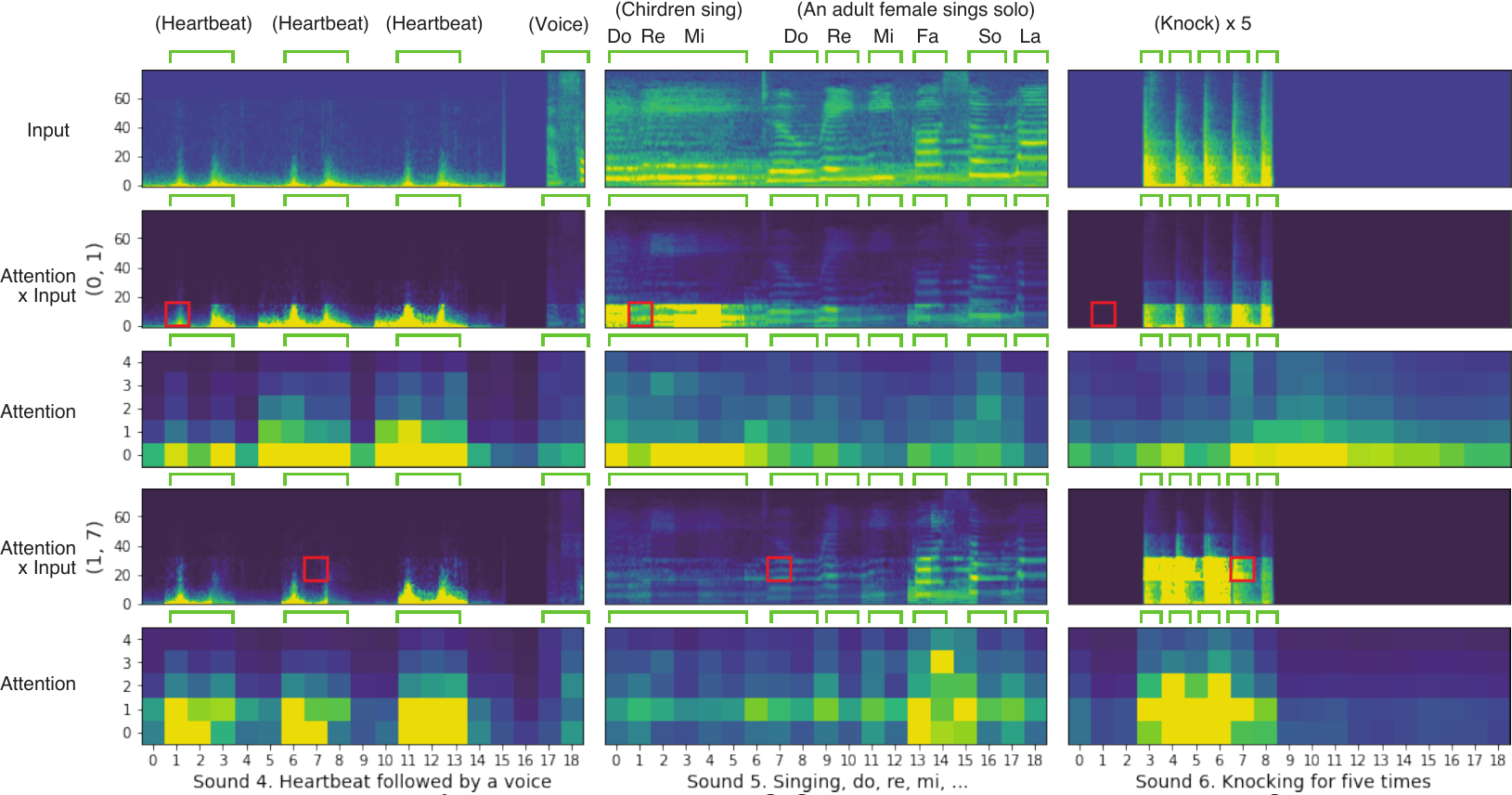}
\vspace{-30pt}
}
\vspace{-5pt}
\end{figure*}

We also observe coarse segmentation of similar sounds for the reference points. For example, sound 4 shows that heartbeats are segmented in the self-attention maps of both reference points, less attending to the following voice; sound 5 shows that the first half of children's singing voices are segmented in the first reference point, whereas the latter half of an adult female voice is coarsely segmented in the second reference point.
%These results show that the encoder puts higher weights on the patches of similar sounds, resulting in making the latent representation focus on these sounds.

\section{Conclusion}
In this paper, we sought to learn audio representation from the input itself as supervision
by using a pretext task of modeling masked spectrogram patches, which we call Masked Spectrogram Modeling (MSM).
To implement MSM, we employed Masked Autoencoders (MAE) with audio spectrogram as input.

We conducted evaluations on the HEAR 2021 NeurIPS Challenge using its benchmark suite across a variety of domains, including speech, environmental sound, and music.
Our models outperformed the HEAR 2021 Challenge results on seven out of 15 tasks (e.g., accuracies of 73.4\% on CREMA-D and 85.8\% on LibriCount) while showing top performances on other tasks where specialized models perform better.
In addition, we investigated design choices of input audio duration and patch size and confirmed that longer duration and finer time resolution with a smaller patch size improve performance.

We also conducted qualitative analyses on various visualizations of outputs from both the MAE encoder and decoder. We observed frequential and temporal structures in the reconstruction results and the self-attention maps, suggesting that the learned representations hold information related to these structures.

While this study does not provide an exhaustive exploration, the quantitative results proved the effectiveness of MSM using MAE, and the qualitative observations suggested useful information in the learned representations.
We believe that the MSM framework holds promising further improvements and applications in the future.

\bibliography{refs}

\newpage
\appendix

\section{Reconstruction Examples of Various MSM-MAE Models}\label{appendix:viz-recon}

\begin{figure*}[htbp]
\floatconts{fig:viz-recon-rand96-512}
{\caption{Reconstructions of the models for short and long input audio duration.}}
{
\subfigure[Examples of the MSM-MAE-96 (input duration: 960 ms).]{\includegraphics[width=\textwidth]{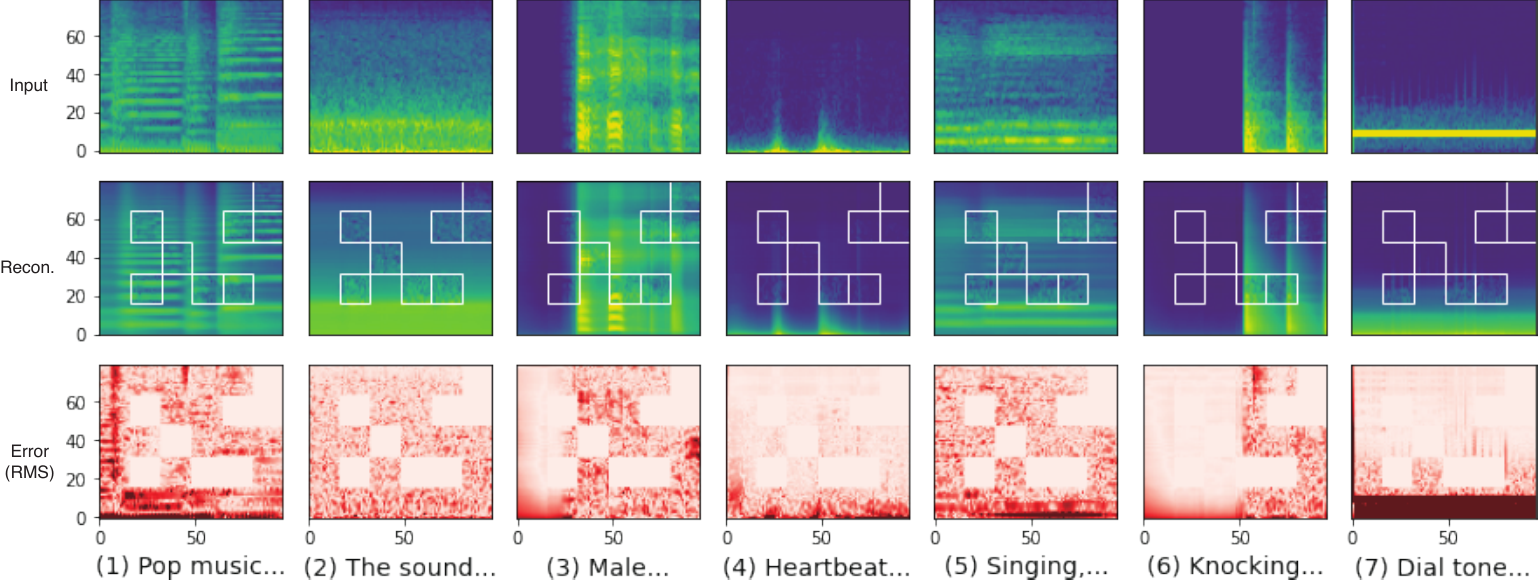}}
\subfigure[Examples of the MSM-MAE-512 (input duration: 5.12 seconds).]{\includegraphics[width=\textwidth]{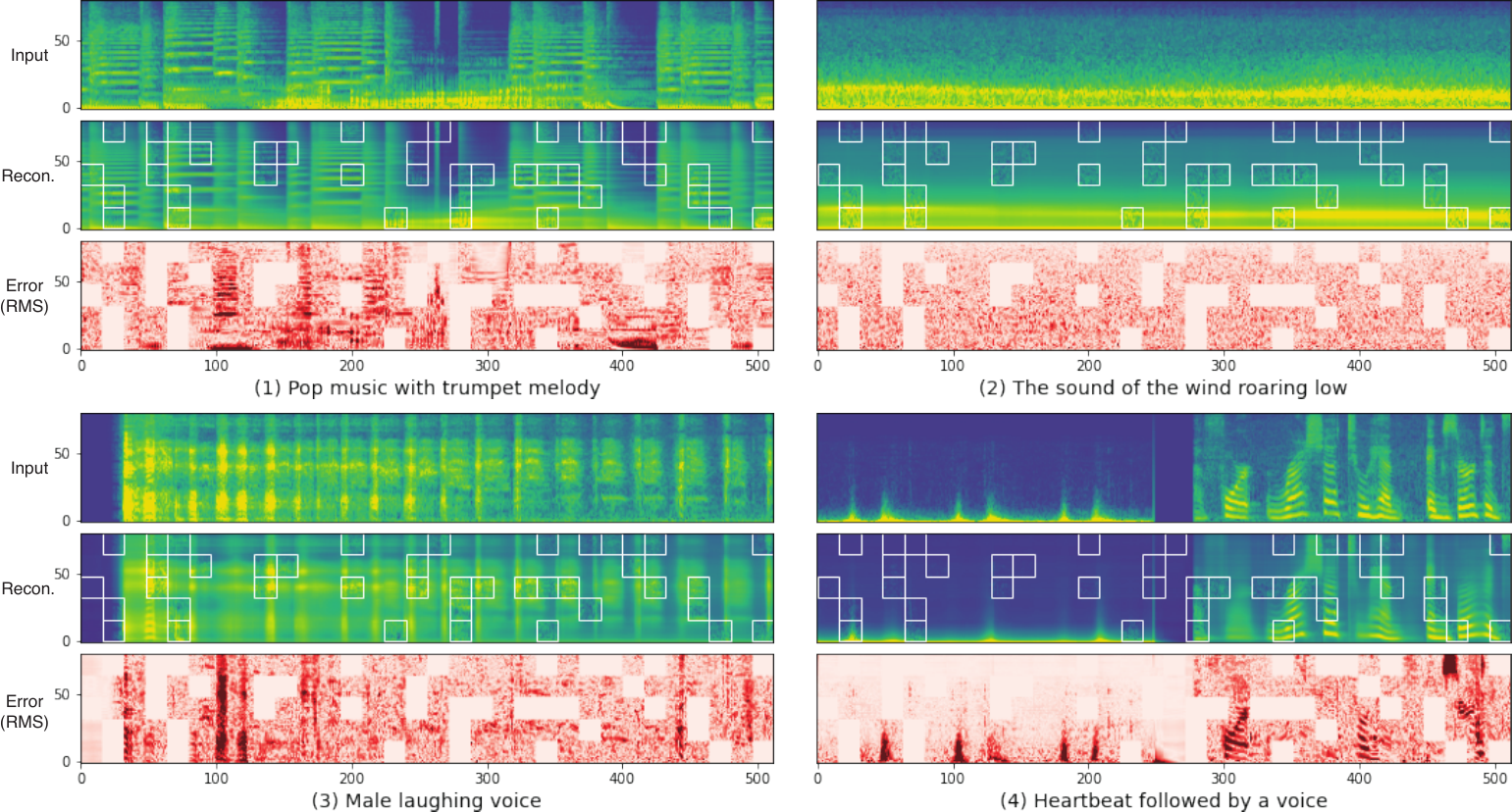}}
}
%\vspace{-10pt}
\end{figure*}

\begin{figure*}[htbp]
\floatconts{fig:viz-recon-rand168-816}
{\caption{Reconstruction examples of the models with various patch sizes.}}
{
\subfigure[Examples of the MSM-MAE-200 (patch size: $16\times 8$).]{\includegraphics[width=\textwidth]{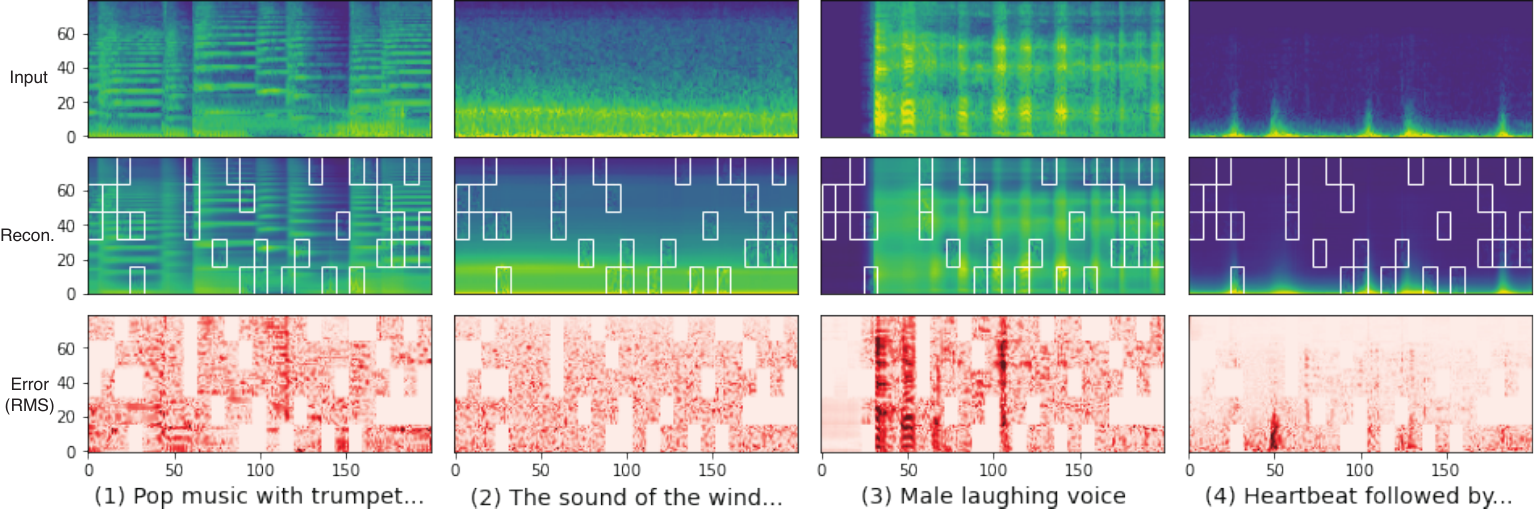}}
\subfigure[Examples of the MSM-MAE-200 (patch size: $16\times 4$).]{\includegraphics[width=\textwidth]{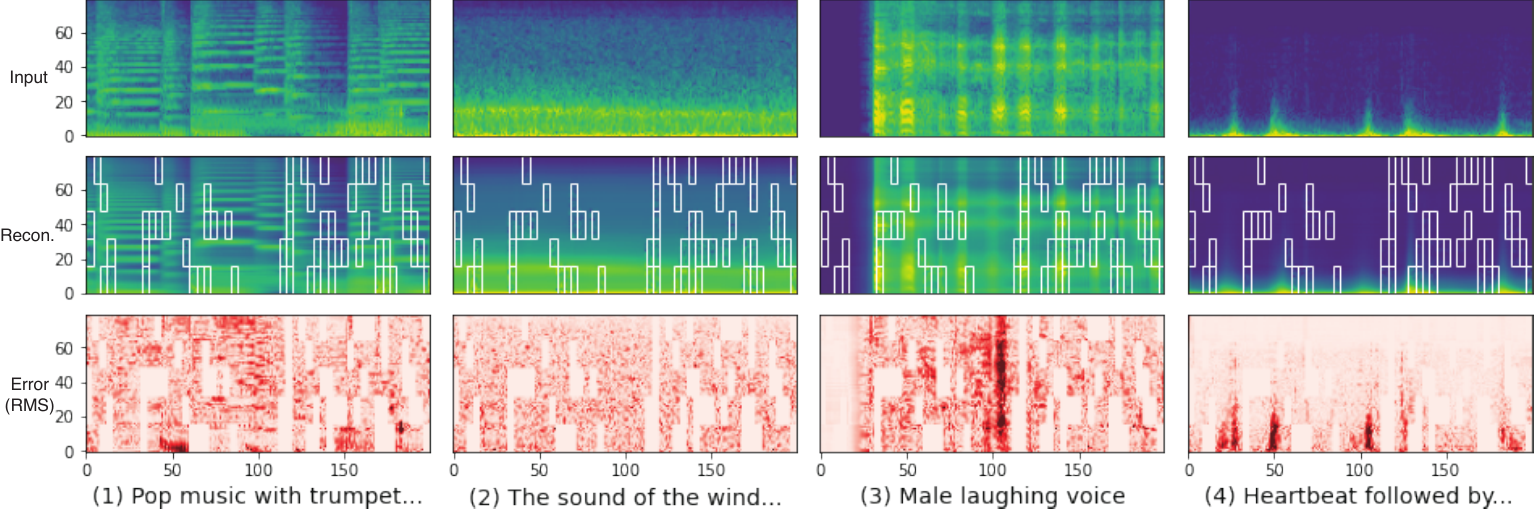}}
\subfigure[Examples of the MSM-MAE-208 (patch size: $8\times 16$).]{\includegraphics[width=\textwidth]{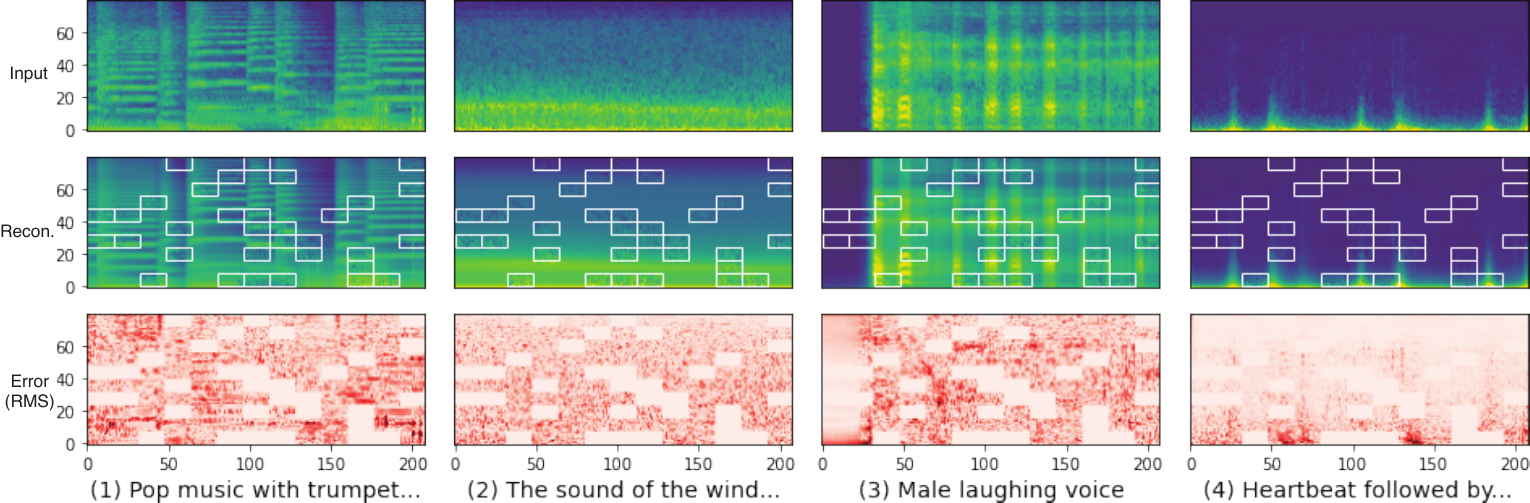}}
}
%\vspace{-10pt}
\end{figure*}

\end{document}